# Chalcogenide optomemristors for multi-factor neuromorphic computation

Syed Ghazi Sarwat[1,†], Timoleon Moraitis[2,#], C David Wright[3] and Harish Bhaskaran[1]

[1]Department of Materials, University of Oxford, Oxford, OX1 3PH Oxford, UK
[†]Present Address: IBM Research- Europe, Säumerstrasse, 8803 Rüschlikon, Switzerland
[2]IBM Research- Europe, Säumerstrasse, 8803 Rüschlikon, Switzerland
[#] Present Address: Huawei - Zurich Research Center, Zurich, Switzerland
[3]School of Engineering, University of Exeter, Exeter, EX4 4QF UK

**Abstract**

Neural processing on devices and circuits is fast becoming a popular approach to emulate biological neural networks. Elaborate CMOS and memristive technologies have been employed to achieve this, including chalcogenide-based in-memory computing concepts. Here we show that nano-scaled films of chalcogenide semiconductors can serve as building-blocks for novel types of neural computations where their tunable electronic and optical properties are jointly exploited. We demonstrate that ultrathin photoactive cavities of Ge-doped Selenide can emulate the computationally powerful non-linear operations of three-factor neo-Hebbian plasticity and the shunting inhibition. We apply this property to solve a maze game through reinforcement learning, as well as a single-neuron solution to the XOR, which is a linearly inseparable problem with point-neurons. Our results point to a new breed of memristors with broad implications for neuromorphic computing.

## Introduction

Efficient neuromorphic sensors and processors have emerged, originally based on CMOS technology[1,2] and more recently using memristive nanodevices, or memristors[3,4]. These are devices that not only can memorize a value in their adjustable physical state but also can use that value to perform computations by modifying an externally applied signal, for example, provide a form of synaptic weighting that depends on the temporal or spatial separations of two or more signals[5–7]. Synapses that utilize the interaction of two distinct signals such as an electrical and an optical signal are an interesting class and have utility in some tasks, viz. a synthetic-framework equivalent to optogenetics[8,9]. Here we demonstrate just such a framework, using germanium selenide ($GeSe_3$) memristive nano-cavity devices that respond to both optical and electrical input signals. We then show that such dual (electrical and optical) control allows neurosynaptic processing, including three-factor synaptic plasticity enabled reinforcement[10] and surprise based learning[11], top-down feedback governed supervised learning[12], and shunting inhibition[13].

Reinforcement learning (RL) is a category of machine learning that is commonly used to learn rewarding strategies. Deep RL, i.e. RL applied to deep neural networks, has resulted in impressive results for artificial intelligence, such as the outperforming of humans in the game of Go[14]. The neural network's synaptic weights in these scenarios are updated based on the interaction between, firstly, a temporal signal, called an eligibility trace (and indicating how far in the past an action was taken at a particular state), and secondly, a possible reward signal if the sequence of actions is successful. Preliminary results on RL using memristive synapses do exist but have relied on hybrid digital-analog approaches, where memristors perform subordinate computational operations, since they only accommodate a single, electrical input signal; in these examples, the actual learning aspect is carried out in digital CMOS[15]. There are additional neural operations such as shunting inhibition that are based on the interaction of multiple signals which enable computations in biological neurons that are not achievable using standard artificial neurons[16–18]. For instance, the XOR logic gate is a common example of a linearly non-separable problem of classification, that requires multiple layers of conventional artificial neurons for its solution. Nevertheless, it has recently been shown that a single biological neuron can solve this problem using dendrites[16]. We demonstrate this too using our chalcogenide nano-cavity devices.

## Crossbar Optomemristors

Our devices are solid-state crossbars (Figure 1A), comprising stacks of thin films of top and bottom electrodes with $GeSe_3$ sandwiched in between (see Section S1-S2). The electrical resistivity of our device is determined by a conductive channel between the top and bottom electrodes, the formation of which is controlled by an electrical field. However, in our devices, resistive switching can also be controlled optically. This optical responsivity introduces an additional control mechanism. Figure 1B illustrates the current-voltage (I-V) characteristics of an $Ag/GeSe_3/Ag$ stack under dark (blue trace) and illumination (red trace) conditions. The high resistance state (HRS) of the device indicates an incomplete conductive channel between electrodes, and the low resistance state (LRS) is indicative of an intact and conductive channel, with these two states being stable (i.e. non-volatile (also see Figure S1.1). However, under optical illumination at 637 nm, the RESET (channel rupture) voltages of our devices shift from the negative bias to the positive bias, i.e. the memristor loses its non-volatility since after SET the device spontaneously RESETs when the applied voltage is removed The threshold voltage ($V_{TH}$) or switching voltage, at which the conductive channel forms in these devices also increases under optical illumination. We find the shift in $V_{TH}$ to be notably significant (~100% relative change) across devices, for modest sub-mW optical power (at 637 nm). When one of the Ag electrodes is replaced with Pt (see Figure 1C), the device ($Pt/GeSe_3/Ag$) spontaneously turns-off (LRS→HRS) when the voltage is ramped below some holding voltage (i.e. becomes volatile). The optical modulation is also observed in such volatile devices, where the device maintains its volatility, while its switching voltage shifts to larger threshold voltages under optical exposure.

Under increasing illumination, we observe the shift in the switching voltage in both the non-volatile ($Ag/GeSe_3/Ag$) and volatile ($Pt/GeSe_3/Ag$) device types to scale proportionally with the intensity of optical illumination (see Figure 1D (for $Ag/GeSe_3/Ag$) and Figure S1.2). Interestingly, optical illumination not only changes the voltages at which the device switches to a higher conductance state, but also induces a zero-voltage current. This behavior is similar to the functioning of a solar-cell and is suggestive of a photovoltaic effect in the devices, which stems from asymmetric Schottky junctions[19]. In Figure 1D, the shifts in the cross-over point $V_O$ (open-circuit voltage, where current is zero) and the negative short-circuit current (at zero-voltage) are plotted as a function of illuminating optical intensity. The values scale logarithmically with optical intensity, with the cross-over point undergoing a shift by 455 mV for an optical intensity of 1 mW. For a device that operates under a photovoltaic mode,[19,20] the direction of the short-circuit changes with the device polarity, which is indeed observed in our devices. This is illustrated in Figure 1E where the photocurrent (short-circuit current) at zero voltage for a $Pt/GeSe_3/Ag$ device is plotted as a function of the illuminated optical intensity under differing polarities.

All the layers in our devices are optically thin (the electrodes are typically sub-50 nm thick, the chalcogenide layer typically 20 to 100 nm thick). These layers behave as resonating optical nano-cavities[21,22] The cavity design allows us to make devices to selectively interact with different wavelengths, for example from the visible to the infrared, through appropriate layer thickness control. [8]. We modeled such effects using transfer matrix calculations (see section S3), with exemplar results shown in Figure 1F where an $Ag/GeSe_3/Ag$ device is designed to work the blue, green, red, and infrared regions of the spectrum. Light absorption with high-quality factors and near-unity absorption coefficients are possible by simply varying the thicknesses of the constituting layers.

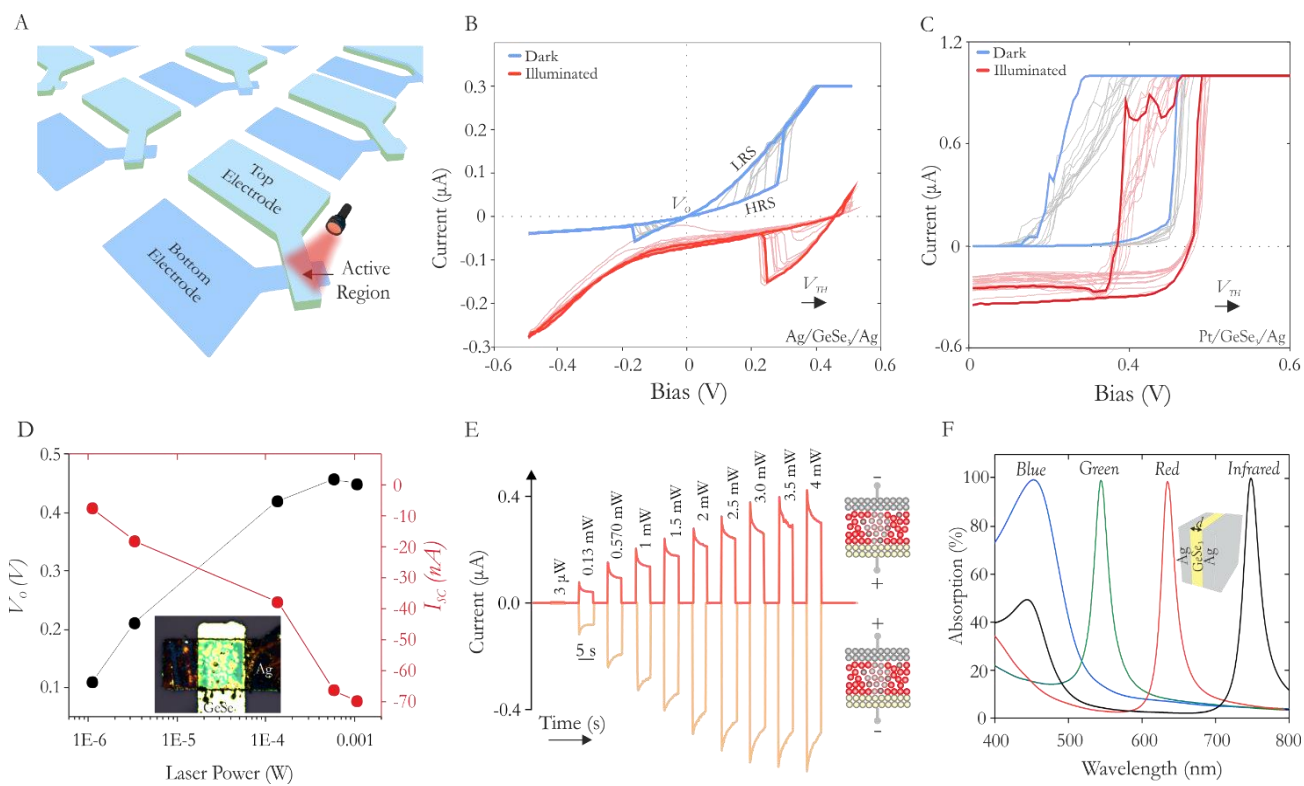

**Figure 1: Optomemristor.** (A) An illustration of crossbar devices. (B) Current-voltage traces of an Ag/GeSe$_3$/Ag non-volatile device under dark and under illumination (637 nm/1 mW) (multiple measurements are overlaid). (C) Similar measurements on a Pt/GeSe$_3$/Ag volatile type switching device. In (B) and (C) the saturation in current is a measurement artifact due to the compliance current values set during measurements. (D) Cross-over voltage ($V_o$) and the short circuit current ($I_{sc}$) as a function of the illuminating laser power. Inset is an optical micrograph of the device. (E) Short-circuit current in a device as a function of illumination intensity (637 nm) under positive and negative device polarities. (F) The absorption spectra in Ag/GeSe$_3$/Ag stacks for blue (d = 28 nm), green (d = 51 nm), red (d = 78 nm), and infrared light (d = 103 nm), for varying thicknesses (*d*) of the GeSe$_3$ films.

When the polarity of either device (Pt and Ag electrodes-based stacks) is reversed, i.e. in a Pt electrode based device, the Pt pad becomes voltage source, while Ag pad the sink, we find that the conducting channels are still formed. However, with the inverted polarity, optical illumination causes the shift in the switching voltage towards smaller values, i.e. switching occurs at reduced voltages under optical illumination (Figure S1.2). Importantly, these observations show that the devices can be optically controlled to either switch at smaller or at higher voltages relative to their intrinsic switching voltage. We also observed a variable time delay (latency) for the onset of switching, with the delay being a function of the applied bias, with voltages closer to the switching voltage increasing the spontaneity of switching (see Figure 2A and S5.1). However, for electrical conditions under which that device does not switch, we find that optical stimulation can result in switching. In Figure 2B, for example, we show the response to illumination of a device biased near its switching voltage (in dark conditions). For low-intensity illumination (yellow trace, (0.13 mW), the device undergoes a series of minor switching events before fully switching into the LRS. This effect is similar to short-term plasticity, where a high conductance non-volatile state is realized through a series of intermediate states and is a feature that has been utilized for rehearsal-based machine learning[23]. However, for illumination intensities that can change the switching voltage to below the applied bias voltage (here 0.39 V), the device undergoes spontaneous switching into the LRS (red trace). Such behavior is shown in Figure 2C, where a pulsed threshold illumination spontaneously switches the device. Once the device is switched, it retains its state (LRS), and the device can then be RESET to the HRS by bringing the applied (bias) voltage to zero. It is clear from the results shown in Figures 1 and 2 that our GeSe$_3$ optomemristor devices offer not only the functionalities generally associated with electrical memristors but also a range of additional, advanced functionalities arising from the combination of electrical and optical means.

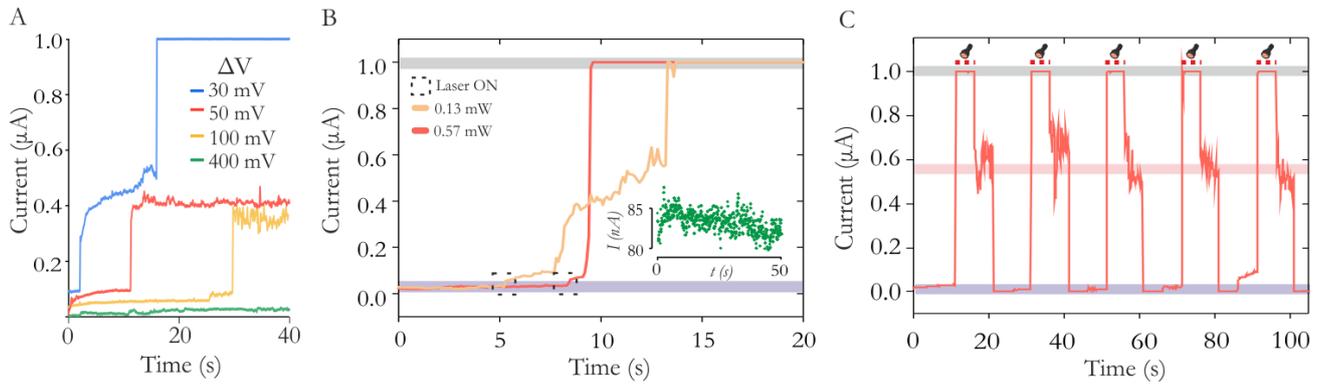

**Figure 2: Switching Dynamics in Volatile Pt/GeSe$_3$/Ag devices.** (A) Stochastic switching processes under dark conditions. ΔV is the difference between the applied constant bias and the device's intrinsic switching voltage. (B) Light-induced switching under a sub-threshold illumination intensity, and at high optical intensity illumination for bias polarity that produces positive short-circuit current. Inset is the device current under dark conditions. (C) Switching behavior of the device to pulsed illumination. For threshold illumination, the device spontaneously switches to LRS.

## Neuromorphic Computing

### Three-factor synaptic plasticity

As a first example of the computing capabilities possible using our optomemristor devices, we describe the implementation of three-factor synaptic plasticity, with important applications in reinforcement learning. We utilize the combined effects of the electrical and optical stimulus to modulate the switching behavior of non-volatile Ag/GeSe$_3$/Ag type devices. In Figure 3A, a device is continuously biased at a voltage of 100 mV, below the switching threshold. Electrical pulses that are 500 ns wide (with the rise and decay time of 5 ns) and 400 mV in amplitude are applied to the device under dark conditions. These voltage pulses alone cannot induce switching of the device. Similarly, we also observe that when the device is illuminated but no voltage is applied (blue trace), the device does also not undergo switching (see inset). However, when the device is illuminated and an electrical pulse is then applied, the device spontaneously switches to a conducting state. This is the optomemristive property that we now exploit to deliver a form of neo-Hebbian learning.

Traditionally, computational neuroscience and neuromorphic computing have been focusing on learning algorithms based on Hebbian types of synaptic plasticity. While Hebbian learning can be successful in several unsupervised learning scenarios, and even outperform supervised deep learning schemes in certain cases such as object classification[24–26], recent work has shown that other important biological learning tasks, broadly those governing aspects of behavioral learning (e.g. motor control, guidance, and navigation) and synaptic consolidation and tagging require the influence of a third factor, separate from the activations of the two local neurons[10,27,28]. Such neo-Hebbian learning requires three-factor plasticity rules[10,29] that involve a delayed third signal, such as the reward, as a key factor in the learning process.

Navigation is one important area of learning in which three-factor plasticity is thought to play a key role. For instance, a neuronal cell representing a spatial location of a rodent in a maze, i.e. a place cell, and an action cell representing a particular action that the rodent may take, will strengthen the synaptic connection between them if the rodent takes that action frequently when it finds itself in that location: as a result, the rodent would learn habitually to take that action in that location. This co-activation of the place and action cells is a form of three-factor, neo-Hebbian learning. In this case, the effect manifests itself as a so-called eligibility flag[10,27,29,30] that makes the pair's (i..e. action and place neuron) synapse eligible for an update if, and only if, a reward signal is provided within a limited time window. We demonstrate this via an example game where a rodent navigates in a maze to find cheese and avoid traps (see Figure 3B). A piece of cheese is hidden in one corner of the maze, while the other corners have traps that cause the rodent to be placed back to its starting position. The rodent must learn the appropriate action to associate with each position, which will lead it consistently to the cheese. The action can be one of four (move north, east, south, or west), and the mapping between a place and an action is represented in a neural network, connecting place cells to action cells. For a particular place cell, the most strongly connected action is chosen. The connections, i.e. synapses, are implemented by our optomemristors.

In the implementation of the above maze, we measure the conductances of the HRS and LRS states of individual devices, before, during, and after their electrical and optical illumination. The extracted conductances are then used as synaptic weights and as updates, in performing the RL simulation on a standard computer. In this simulation, initially, the neural network comprising the synapses is untrained, and all synaptic weights are set to zero., i.e. devices are reset to their low conductance state. The training happens as follows (see Figure 3C). The rodent explores the maze by taking random actions. Every time an action is taken, the eligibility flag is raised at the corresponding synapse, which in our case is by illuminating the device. This eligibility trace can have arbitrary profiles, such as exponentially decaying or stepped. The stepped waveform is typically used in the synthetic implementations of RL, including in this work[10,31,32]. In our example, we arbitrarily choose the flag to remain raised for three-time steps but allows the expectation for a reward for all possible trials including when the longest path to the cheese is taken in this particular example. If the rodent finds the cheese, an electrical pulse representing a global reward signal is given to all synapses, both the inactivated and those activated during the exploration trial. But, only the last three place-action pairs, if they lead to the reward are potentiated to their high conductance state. Through this process, the rodent successfully learns the correct synaptic weights between the place and action cells that guide the rodent to the cheese in future. In Figure 3D we plot a quality (Q)-table of the learned network, which maps the favorable actions when in given states through the conductance states of the synapses. For example, when in the maze center, the rodent will preferably move north (up direction), since the corresponding synapse has the highest conductance. Overall, this practical example demonstrates synaptic devices with in-situ three-factor plasticity, and their application in an RL task, made possible by the multi-factorial, i.e. optical and electrical, response properties of these memristors.

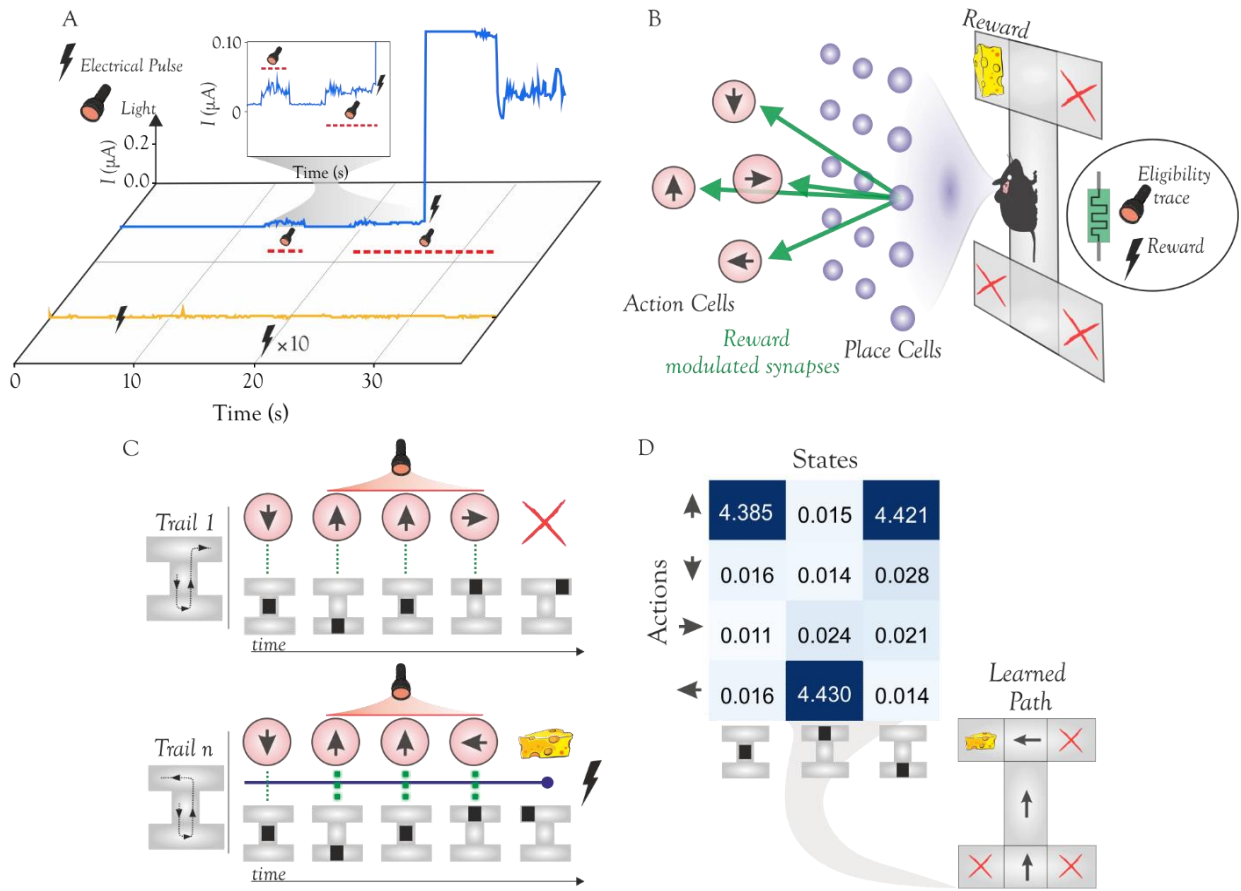

**Figure 3: Emulation of three-factor synaptic plasticity and reinforcement learning.** (A) Mixed mode behavior of a non-volatile Ag/GeSe$_3$/Ag device. In the absence of light (yellow trace), electrical pulses applied to the device do not induce a switching event. Under illumination (blue trace) however, electrical pulses can trigger HRS to LRS switching. (B) Sketch of a rodent in a maze. Place cells represent the rodent's location in the maze, such that, at each location, one unique place cell is active. Each action cell represents one of four movement directions, and each location triggers one of the four movements. Initially, all synaptic weights equal zero, and through the exploration of the maze and reinforcement learning the rodent learns the weights that enable correct navigation from the initial position to the cheese reward. Each synapse and its weight here are emulated by a non-volatile type memristive device and its conductance. Reinforcement learning emerges through a three-factor synaptic plasticity rule. The rule involves an eligibility flag, which in our case is the illumination of the corresponding memristor, and a reward applied as an electrical signal sent to all memristors. (C) Example trials during the rodent's training. Each time the rodent moves, an eligibility flag (optical signal) is raised at the synapses of the corresponding place and action cell (red trace). The three eligible synaptic weights are not updated by the optical flag alone, e.g. in an unsuccessful trial (top sequence). A successful trial provides the electrical reward that potentiates the eligible synapses (bottom sequence). (D) Results of training. In the learned weight matrix of the neural network, the electrical conductance (in µS) of the memristive synapses maps each place to an action. This learned mapping corresponds to the correct path to the cheese (inset).

## Shunting Inhibition

We now demonstrate the second example of single-device multi-factor neuromorphic computation that is enabled by our optomemristors. For this demonstration, we use the volatile devices to emulate single-device dendrites with shunting inhibition, one of the three fundamental types of connectivity between biological neurons[13,33–35]. We then use these shunting dendrites to demonstrate a single-neuron implementation of an XOR logic gate. The XOR is a textbook example of a classification problem that in ANNs requires a network rather than a single neuron for its solution[17,18,36]. Biological neurons however can implement XOR by virtue of nonlinear operations in their dendrites[16]. We here use our optomemristor devices to reproduce this increase in computational power brought about by dendrites. In Figure 4A, the relevant electro-optical control of a Pt/GeSe$_3$/Ag device is illustrated: when illuminated the device's conductance drops, due to the generation of a negative photocurrent. For these conditions, even when electrical pulses are applied the device is restricted from undergoing filamentary switching. Under dark conditions, however, the device spontaneously switches from HRS to LRS, when an electrical pulse is applied. This is the property that we exploit.

In biology, shunting inhibition describes the following communication mechanism between neurons[13]. Excitatory input to a neuron causes an excitatory postsynaptic potential (EPSP) that propagates through the dendrite. In certain cases, a shunting inhibitory synapse is attached to the same dendrite, but more proximally to the neuron's soma. If the neuron receives input also from the shunting synapse at the same time as an excitatory input, then the EPSP is shunted, i.e. canceled, or attenuated (see Figure 4B). Our memristive dendrite receives excitatory input as an electrical signal and shunting inhibitory input as optical illumination. An electrical pulse causes current to pass through the device, but coincident illumination inhibits (i.e. shunts) the current, thus mimicking shunting inhibition (see Figure 4B). Notably, illumination alone does not increase the device conductance, representing the shunting-only effect of this inhibition. Thus, a single memristor device is able to emulate a dendrite with two adjoining synapses, namely an electrical excitatory one, and an optical inhibitory one.

We then show how a neuron with two such dendritic devices can implement a solution to the XOR (Figure 4C) 'problem'. Here the neuron Z receives input from two such memristive dendrites and produces the logic output. Dendrite 1 has two inputs: Neuron X sending a synaptic electrical excitatory input and Neuron Y an inhibitory optical one. On the other hand, for dendrite 2, Neuron Y sends an excitatory electrical input and Neuron X send it the optical inhibitory one. The effects of shunting inhibition on each dendrite ensure that if both X and Y are active, the postsynaptic potentials on both dendrites are attenuated. However, if exclusively X or Y is active, then a postsynaptic potential propagates through one of the two dendrites and activates the neuron Z. The graphical representation of the simulated outputs is illustrated in Figure 4D. The dark circles highlight the combinations of inputs (X,Y) that activate the output neuron; for other combinations the neuron is inactive. The green XY plane represents the neuron's threshold, above which the neuron outputs 1. Markedly, the proposed device connectivity realizes an XOR logic gate within a single neuron, by exploiting the added nonlinearity of shunting inhibition at its dendrites. Interestingly, other single-neuron solutions to XOR have been previously hypothesized and considered theoretically possible in biological neurons, contrary to ANNs, due to dendritic nonlinear computations[12,36,37]. One such solution was very recently confirmed experimentally in the human brain[16]. We note that the standard neural net implementation of XOR requires multilayer perceptron with at least ten transistors and six memristors[18,38–40]. Thus, our device represents the increased effectiveness of neuromorphic computing compared to conventional ANNs, and its potential to contribute to realizing these novel applications.

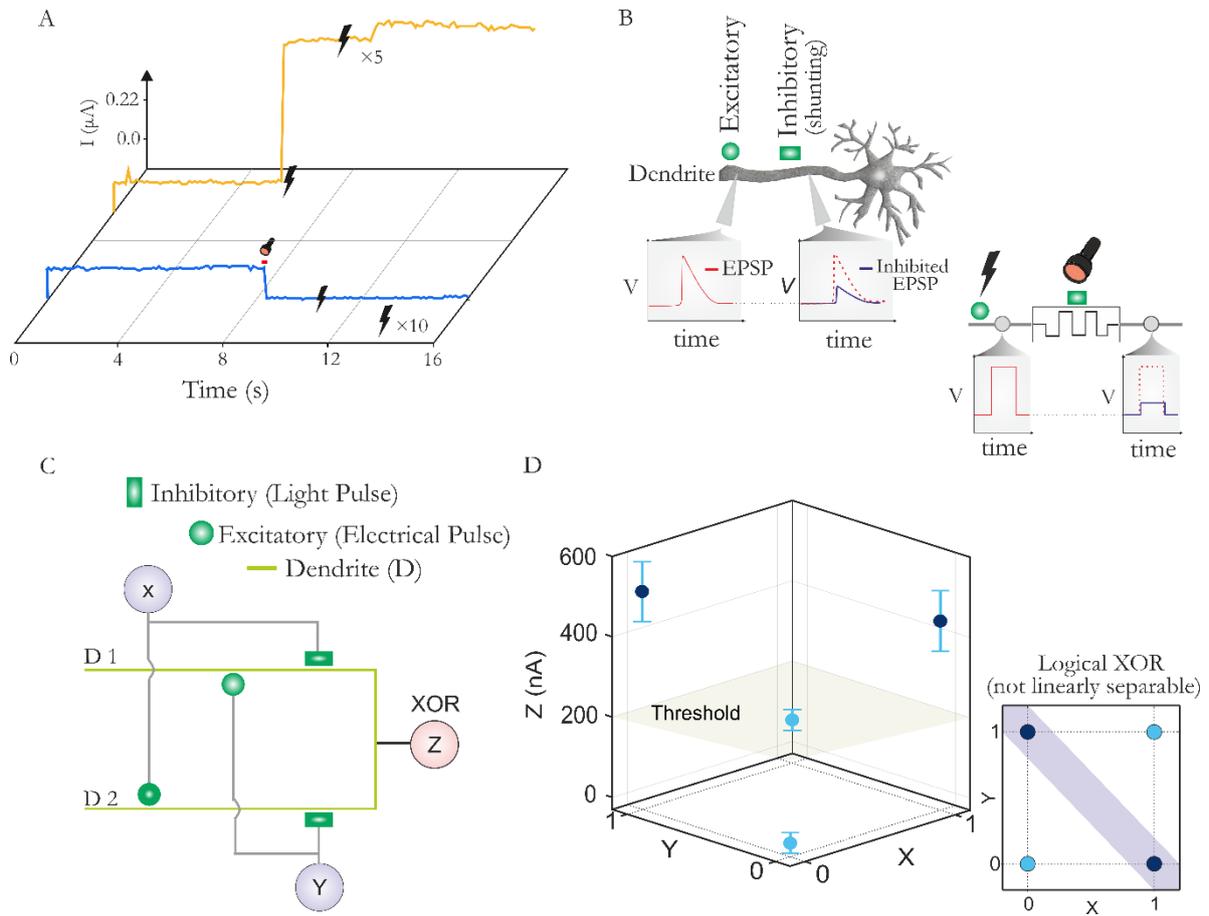

**Figure 4: Shunting inhibition and single-neuron implementation of XOR.** (A) A Volatile Pt/GeSe$_3$/Ag device under a mixed mode operation. In the absence of light (yellow trace), electrical pulses applied to the device induce a switching event, while the presence of light inhibits device switching. (B) Sketch of a biological neuron, with a dendrite possessing an excitatory synapse and, proximally to the neuron's soma, a synapse capable of shunting inhibition. Each excitatory input generates an EPSP (red trace) that propagates, but its effect is gated by inhibitory inputs. An input to the shunting inhibitory synapse attenuates the EPSP (blue trace), but in the absence of excitation, it has no effect (not shown). A volatile type memristive device emulates this dendrite, with the excitatory and shunting inhibitory inputs applied by electrical and optical stimulation respectively. (C) A neuron (Z) comprising two memristive dendrites (D1 and D2) that receive inputs from neurons X and Y. Each dendrite has an excitatory and a shunting input synapse. (D) Results of Z's output for different input pairs. Owing to the memristive dendrites, neuron Z realizes an XOR gate, a function that is impossible for a single layer of point neurons. The green plane defines the activation threshold of Z.

### Device Switching Mechanisms

Overall, we explored seven electrode/GeSe$_3$ material combinations for optomemristive behavior (see supplementary information section S1). Based on the nature of resistive switching (see Figure S1.1) - which was either non-volatile (Ag/GeSe$_3$/Ag devices), volatile (Pt/GeSe$_3$/Ag devices), or absent (all other combinations) - we infer that the switching process involves a mobile element, namely Ag and that the (non) volatility is dependent on the electrode material; ruling-out switching to involve electron instability and oxygen vacancies, effects commonly reported in Ge rich composition of Se based glasses[41,42]. On lateral devices (Ag/GeSe$_3$), we observed that under an applied voltage, the switching event is preceded by the formation of a dendritic structure or the filament (see Figure S1.3). In such devices, the solid-electrolyte was observed to be no longer a uniform thin film structure after sputter-deposition. Instead, it comprised of segregated globule-like structures (likely particles[43] of Ag) that were uniformly embedded in the surface and volume of the GeSe$_3$ matrix. Such nanostructures may exist due to the spontaneity of silver dissolution in chalcogenide glasses[44–47](see Figure S1.6) and were also found to be depleted in the filament's proximity (see Figure S1.5). An elemental spectroscopy map (using energy dispersive X-ray) shows that the filament is rich with constituent elements, including Ag. Furthermore, the molecular make-up of the filament is observed using Raman spectroscopy to be uniquely different from the matrix and bare GeSe$_3$ film (see Figure S1.4). Characteristic peaks of the vibrational modes of Se and Ge−Se are quenched near the filament, complementing the EDX profiles in suggesting that the filament is a multi-component structural unit. We also performed transmission electron microscopy studies. The diffraction patterns showed the GeSe$_3$ film matrix to be amorphous, while the filament to be crystalline. As additional proof of filamentation, we tested devices of differing areas. When in the SET state, the filament dominates the resistance of the device; thus while the HRS should decrease with the device area, the LRS is expected to show no scaling, which is precisely what we observe (see Figure S1.7). These observations combined with the optoelectronic measurements which showed an increased charge collection (larger photocurrent) in the LRS states of devices indicate filamentary behavior as the primary mechanism to describe the observed switching effects. The observed dependency of filament's stability (volatility) on the electrode material is likely dictated by minimization of the interfacial energies between the filamenting material, the dielectric, and the electrode, effects that have been observed in other memristors[48]. The observations so far, collectively suggest that the mode of optical tunability in the devices studied here is photovoltaic and governed not only by simple electromigration of a conductive filament from a host electrode but also by the electric field-driven re-arrangement and precipitation of the already dissolved electrically conductive nanostructures at the electrodes. Such effects have been shown to induce polarity independent filamentation in other material systems, including Ag/ZrO$_2$, Ag/TaO$_5$, Ag/ZnS stacks[48–50].

## Conclusions

In conclusion, we have described a novel framework using GeSe$_3$ devices with silver ions acting as a memristive element that is both electrically and optically active. Such an "optomemristor" is shown to be configurable as both volatile and non-volatile, governed by the choice of the electrodes. Our extensive characterization of these electrodes indicates that it is the movement of Ag ions in the GeSe$_3$ matrix that enables these effects. We then characterize the unique opto-electronic features of these devices and then exploit them for functionalities which enable new and improved computations, such as three-factor plasticity and shunting inhibition neuromorphic applications that rely on and exploit multi-signal interactions. Within the scope of an optical and electrical stimulus, these devices can also enable biomimetic retinal vision, where they can emulate both spiking photodetection and retina's cunning feedback machinery for intensity control. All in all, these devices expand on the use cases of memristors for emerging neuromorphic computing applications, as well as provide a test vehicle for experimenting ideas in neuroscience.

## Methods

**Device fabrication**: Films were sputter-deposited directly on thermally grown 300 nm $SiO_2$ wafers (IDB Technology, UK). Substrates were first cleaned for 10–15 min in acetone under ultrasonic agitation, rinsed in isopropanol, and dried with pressurized nitrogen. The bottom electrode of the cross-bar devices was then patterned using standard photolithography (positive resist-S1813: exposed for 14 s, baked at 120 °C, and developed for 45 s in MF319 developer). Reactive ion etching was carried out to embed the electrodes in the oxide. Ta (16 nm) was deposited as an adhesive layer in a Nordiko sputtering system: working pressure of 9.6 µTorr, 44.5 sccm (standard cubic centimeters per minute) Ar, and 120 W RF. Bottom electrode was then subsequently deposited in the same sputtering system: with a typical working pressure of 3.5 µTorr, 11.5 sccm Ar, and 40 W RF, without breaking the vacuum. Following lift-off in acetone with mild ultrasonic agitation, the top electrodes were patterned using the same photolithography procedure. $GeSe_3$ deposition was then carried out from a solid target (Testbourne, UK): working pressure of 3.5 µTorr, 11.5 sccm Ar, and 30 W RF. Without breaking the vacuum, top was then sputter-deposited (Testbourne, UK): at unique sputtering conditions. Lift-off was carried out in acetone: 65 °C for 8 h. For nano-gap devices, graphene was patterned using electron beam lithography and the nano-gaps were produced using feedback-controlled electroburning. A self-alignment approach described in Nano Letters 2017, 17, 6, 3688–3693 was used for deposition of GST.

**Electrical and optical characterization**: Electrical measurements were carried out using a Keithely 2614 B sourcemeter, Tektronix AFG000C pulse generator, and Teledyne Lecroy WaveSurfer Oscilloscope. The devices were illuminated using a custom-built probe station with a Gaussian beam spot size of 20µm for 637nm laser. Fiber-coupled lasers were used (Thorlabs) for illumination. The devices were wire-bonded using Al/Si wires to a custom-built printed circuit board, which in turn was connected to the measuring units using 50Ω coaxial and SMA cables. All measurements were computerized using custom-built LabVIEW codes. Reflectivity measurements were performed on a custom-built microscope setup. The reflection spectra were simulated using the transfer matrix method, adopted on custom-built MATLAB codes. The refractive index data of the for simulations were experimentally derived using a J.A.Woollam ellipsometer, whereas for Ta and Pt using the existing literature. For the simulations described in the main-text, experimental data on the device conductance under, during, and after optical and electrical pulsing was extracted from a few crossbar cells on the same chip

## Acknowledgments

S.G.S acknowledges the Felix Scholarship which supported his DPhil studies at the University of Oxford, and discussions with Abu Sebastian. We acknowledge the support of Dr. Ian Griffiths and Dr. Kerstin Jurkschat of the University of Oxford for their support with TEM. This work was supported by EPSRC grant numbers EP/R001677/1, EP/M015173/1 and EP/J018694/1 and the John Fell Fund.

## Competing financial interests

The authors declare no competing financial interests. HB holds shares and serves on the board of Bodle Technologies Ltd. And Salience Labs Ltd.

## Supporting Information

X- ray diffraction, Raman Spectroscopy, Energy dispersive X-ray, Current-Voltage graphs, Optical and Scanning electron micrographs, Transfer Matrix Calculations, Spiking Photodetector operation.